\documentstyle[12pt]{article}

\begin{document}
\title{\bf Quantum radiation from a partially reflecting moving mirror}
\author{NISTOR NICOLAEVICI\\
\it The West University of Timi\c soara,\\
\it V. P\^arvan Ave. 4, RO-1900 Timi\c soar, Romania}
\maketitle
\begin{abstract}
We consider the quantum radiation from a partially reflecting moving mirror
for the massless scalar field in $1+1$ Minkowski space. Partial
reflectivity is achieved by localizing a $\delta$-type potential at the mirror's 
position. The radiated flux is exactly obtained for arbitrary motions 
as an integral functional of the mirror's past trajectory. Partial 
reflectivity corrections to the perfect mirror result are discussed.
\end{abstract}

\section{Introduction}\label{one}

One of the most interesting prediction of the quantum field theory 
is the radiation produced by moving mirrors. A mirror can be practically 
seen as a barrier potential in the vicinity of a certain spatial 
boundary. Moving mirrors effects can be placed thus in the larger context of 
that produced in exterior classical fields. Essentially, the phenomenon 
expresses the fact that in presence of mirror-like boundaries in (non-uniform) 
motion real quanta will be excited out from vacuum fluctuations.

For perfect mirrors, situation enjoys a certain technical definiteness. 
As infinite barriers are involved, the problem basically relies in 
assuring the quantum field to vanish on the corresponding boundaries. There 
are a lot of mathematical methods to deal with such problems, and 
a wealth of literature \cite{anderson} following this line can be found. 
(Perfect reflectivity seems also to have been the preferred choice for 
discussing a 
series of most profound aspects of the phenomenon: the connection with the 
Hawking effect \cite{birrell}-\cite{hotta} and non-trivial situations implied by the possibility of 
generating negative energy fluxes \cite{davies1}-\cite{ford2}). As 
concerns partial reflectivity, results 
are comparatively less numerous. Two main directions of research have to be 
mentioned here. One is that followed by Jaekel and Reynaud, based on the 
method developed in Refs. \cite{jaekel1, jaekel2}. Reflectivity properties of 
the 
mirror are described within a scattering approach, and the radiation is 
derived from the analysis of an $S$ matrix depending on the mirror's 
trajectory. The method has a large generality, but it is associated with a 
major 
drawback: by its very nature it is ill suited for an accurate evaluation of 
the radiated flux density, as a $local$ quantity, at arbitrary times. 
In addition, 
the analysis is carried out in frequency space, obscuring somewhat the 
trajectory dependence in coordinate space. Another direction is that 
of Barton et 
al. \cite{barton1}-\cite{barton3} (see also Ref. \cite {gutig}). 
They proceed in the more conventional way of constructing the field 
operator 
via its modes expansion, tacking into account explicitly dielectric properties 
of the mirror. As a common feature, however, their results are systematically 
obtain within the framework of perturbation theory, in first order in the 
mirror's velocity.

To our best knowledge, the only calculation allowing an exact result 
for the radiated flux is that of Frolov and Singh \cite{frolov} for 
spherical mirrors (in $D$-dimensional flat space) expanding with a 
constant acceleration. 
Semitransparency is realized by means of a $\delta$-like potential. Our 
intention here to provide a two dimensional semitransparent mirror model based 
on a similar idea, allowing an exact treatment for arbitrary motions.

We shall focus on the massless scalar field. The essential piece in our 
investigation 
is the construction of the (in) modes defining the field operator in Heisenberg 
image. We obtain them by introducing a set of space time dependent reflection 
and transmission coefficients into those of the Fulling and Davies model 
for perfect reflectors \cite{fulling}. Matching conditions on the mirror's 
worldline
force the coefficients to respect a first order differential 
equation, which determine them as integral functionals of the mirror's past 
trajectory. It turns out that calculations can be exactly carried out 
for the Wightman function and the renormalized energy-momentum tensor (we 
restrict here to the canonical expression). 
Qualitatively, one finds that the radiated flux results as history dependent 
quantity, with past dependence extending in principle to minus 
infinity. Contributions of past motions appear exponentially 
damped with
the
proper time interval up to the emission moment, with the barrier strength
as the proportionality factor. 
The perfect mirror result is recovered in the limit of infinite barrier 
strengths.

The paper is organized as follows. In Sec. \ref{two} we construct the
quantum
field. In Sec. \ref{three} we obtain the in-vacuum Wightman function and
the
renormalized energy-momentum tensor. A brief discussion of the near perfect
mirror approximation (large barrier strengths or slowly varying motions) is
also given. We end in Sec. \ref{five} by drawing a parallel with extended
charges in classical electrodynamics, relying on the history dependence of
the emitted flux. Technical matters are referred to the Appendices.

Natural units $\hbar=c=1$ are used throughout the paper. The metric tensor
is
$g_{tt}=-g_{xx}=1$.

\section {The quantum field}\label{two}

Let $(t,x)$ denote the coordinates in 1+1 Minkowski space time and
$x(t)$ denote the trajectory of the mirror. We write the barrier potential
as
\begin{eqnarray}
V(t,x)=\frac{a}{\gamma (t)}\delta (x-x(t)),
\quad \gamma=\biggl(1-\biggl(\frac{dx}{dt}\biggl)^2\biggl)^{-\frac{1}{2}},
\label{1.3}
\end{eqnarray}
where $\delta$ is the Dirac distribution and $a$ a positive constant
corresponding to the barrier strength. The $\gamma^{-1}$ contraction factor 
assures relativistic covariance of Eq. (\ref{1.4}) below.
We proceed to construct the in modes for the field $\varphi(t,x)$ obeying
\begin{eqnarray}
(\Box+V(t,x))\varphi(t,x)=0.
\label{1.4}
\end{eqnarray}

We divide first the Minkowski plane into the left ($L$) and right ($R$)
regions, corresponding to points $(t,x< x(t))$ and $(t,x> x(t))$,
respectively.
Let us write the in modes for the perfect mirror \cite{birrell} as
\begin{eqnarray}
U_\omega^L(u,v)=e^{-i\omega u}-e^{-i\omega f(v)}, \quad
U_\omega^R(u,v)=0,
\hspace*{44pt}
\label{uperf}\\
V_\omega^R(u,v)=e^{-i\omega v}-e^{-i\omega g(u)}, \quad
V_\omega^L(u,v)=0, \quad \omega>0,
\label{vperf}
\end{eqnarray}
where $L,R$ superscripts refer to the definition domains and
null coordinates were used ($u=t-x$, $v=t+x$). Functions $f$, $g$ are chosen
such that identities
\begin{eqnarray}
f(v)=u,
\label{fv}\\
g(u)=v,
\label{gu}
\end{eqnarray}
are satisfied when $(u,v)$ moves along the trajectory.

Consider, for example,
the $U_\omega^L$ solution. $e^{-i\omega f(v)}$ represents the totally
reflected component of the incident right-moving wave $e^{-i\omega u}$.
Partial reflectivity requests introduction of a reflection factor, along
with a
transmitted component in $R$. We modify thus expressions (\ref{uperf}) as
\begin{eqnarray}
U_\omega^L(u,v)&=&e^{-i\omega u}-{\cal R}_\omega^L(v) \,e^{-i\omega f(v)},
\label{1.7} \\
U_\omega^R(u,v)&=&{\cal T}_\omega^R(u)\, e^{-i\omega u}.
\label{1.7*}
\end{eqnarray}
For $V_\omega^{L,R}$ we similarly set
\begin{eqnarray}
V_\omega^R(u,v)&=&e^{-i\omega v}-{\cal R}_\omega^R(u) \,e^{-i\omega g(u)},
\label{1.8}\\
V_\omega^L(u,v)&=&{\cal T}_\omega^L(v)\, e^{-i\omega v}.
\label{1.8*}
\end{eqnarray}
The physical significance of ${\cal R}_\omega^{R,L}$, ${\cal
T}_\omega^{R,L}$
coefficients is clear. For points inside the $L$, $R$ regions, the
$u$, $v$ dependence in Eqs. (\ref{1.7})-(\ref{1.8*}) automatically assures
\begin{eqnarray}
\Box U_\omega^{L,R}=\Box V_\omega^{L,R}=0.
\label{free}
\end{eqnarray}
At the $L$-$R$ boundary, for each pair of
solutions $(\varphi_\omega^L$, $\varphi_\omega^R)_{\varphi=U,V}$ we
impose the continuity condition
\begin{eqnarray}
\varphi_\omega^L=\varphi_\omega^R,
\label{cont1}
\end{eqnarray}
and the matching of the first order derivatives, in conformity with
Eqs. (\ref{1.3}), (\ref{1.4}). This is most conveniently obtained by
considering the relativistic
generalization of the stationary case $x(t)=$const. One finds
\begin{eqnarray}
\epsilon_{\mu\nu}U^\mu\partial^\nu\!\varphi_\omega^R-
\epsilon_{\mu\nu}U^\mu\partial^\nu\!\varphi_\omega^L
+a\,\varphi_\omega^{R/L}=0,
\label{cont2}
\end{eqnarray}
where $\epsilon_{\mu\nu}$ is the totally antisymmetric tensor in two
dimensions $(\epsilon_{tx}=1)$ and $U^\mu$ is the mirror's two velocity
($U^\mu U_\mu=1$). Equation (\ref{cont1}) implies
\begin{eqnarray}
{\cal R}_\omega^L+{\cal T}_\omega^R=
{\cal R}_\omega^R+{\cal T}_\omega^L=1,
\label{1.9}
\end{eqnarray}
and Eq. (\ref{cont2})
\begin{eqnarray}
\frac{d{\cal T}_\omega^R}{du}-\frac{d{\cal R}_\omega^L}{dv}
+i\omega\left(1+\frac{df}{dv}\right){\cal R}_\omega^L+a\gamma\, {\cal
T}_\omega^R=0,
\label{1.10}\\
\frac{d{\cal T}_\omega^L}{dv}-\frac{d{\cal R}_\omega^R}{du}
+i\omega\left(1+\frac{dg}{du}\right){\cal R}_\omega^R+a\gamma\, {\cal
T}_\omega^L=0,
\label{1.11}
\end{eqnarray}
where all quantities refer to a given point on the trajectory.

Let $\tau$ denote the proper time of the mirror,
and let us regard 
${\cal R}$, ${\cal T}$ coefficients as functions of $\tau$ via
$u=u(\tau)$,
$v=v(\tau)$. Eliminating ${\cal T}_\omega^{R,L}$ in favour of ${\cal R}_
\omega^{L,R}$ using Eq. (\ref{1.9}), one finds that Eqs. (\ref{1.10}) and 
(\ref{1.11}) combined lead to the 
following differential equation for ${\cal
R}^{L,R}_\omega$
\begin{eqnarray}
\frac{d}{d\tau}
{\cal R}^{L,R}_\omega(\tau)+\left(-i\omega D_\beta^{\pm}(\tau)+\frac{a}{2}
\right){\cal R}_\omega^{L,R}(\tau)=
\frac{a}{2},
\label{1.12}
\end{eqnarray}
where we introduced the Doppler factors ($\beta=dx/dt$)
\begin{eqnarray}
D_\beta^{\pm}(\tau)=\sqrt{\frac{1\pm\beta(\tau)}{1\mp\beta(\tau)}}.
\label{1.13}
\end{eqnarray}
Plus and minus signs in $D_\beta^{\pm}$ correspond to $R$, $L$,
respectively. For example, for uniform trajectories $\beta=$const the solution 
reads
\begin{eqnarray}
{\cal R}_{\omega}^{L,R}=\frac{a/2}{-i\omega D_\beta^\pm+
a/2},
\label{1.14}
\end{eqnarray}
having a transparent interpretation in terms of Doppler shifts
as observed in the mirror's proper frame.

The general solution can be written
\begin{eqnarray}
{\cal R}_\omega^{L,R}(\tau)=
{\cal R}_\omega^{L,R}(\tau_0)\exp\left (i\omega\delta^{\pm}(\tau,\tau_0)
-a(\tau-\tau_0)\right )
\nonumber\\
+\frac{a}{2}\int_{\tau_0}^{\tau} d\tau^\prime\exp\left
(i\omega\delta^{\pm}(\tau,\tau
^\prime)-\frac{a}{2}(\tau-\tau^\prime)\right ),
\label{1.15}
\end{eqnarray}
with $\tau_0$ fixed and
\begin{eqnarray}
\delta^{\pm}(\tau_2,\tau_1)=\int_{\tau_1}^{\tau_2}
d\tau^\prime D_\beta^{\pm}(\tau^\prime).
\label{1.15***}
\end{eqnarray}
With definitions (\ref{1.13}) one actually finds for $\delta^\pm$
\begin{eqnarray}
\delta^{+}(\tau_2,\tau_1)=v(\tau_2)-v(\tau_1),
\label{1.15*}\\
\delta^{-}(\tau_2,\tau_1)=u(\tau_2)-u(\tau_1).
\label{1.15**}
\end{eqnarray}
We let now $\tau_0 \rightarrow -\infty$. The first term in Eq. (\ref{1.15})
vanishes\footnote{$\vert {\cal R}_\omega^{L,R}(-\infty)\vert <1$ by the
assumption of uniform velocity in the infinite past, see next.}, hence 
\begin{eqnarray}
{\cal R}_\omega^{L,R}(\tau)=
\frac{a}{2}\int_{-\infty}^{\tau} d\tau^\prime \exp \left
(i\omega\delta^{\pm}(\tau,\tau
^\prime)-\frac{a}{2}(\tau-\tau^\prime)\right).
\label{1.16}
\end{eqnarray}
We take Eq. (\ref{1.16}) as the definition for the reflection coefficients.
Transmission factors follow from unitarity relation Eq.(\ref{1.9}). This
completes the derivation of the in modes.

An observation is appropriate here: by the real part of the exponential 
factors, the coefficients are determined mainly by the motion 
in an interval
\begin{eqnarray}
0<\tau-\tau^\prime \stackrel{<}{\sim} a^{-1}.
\label{effective}
\end{eqnarray}
As an immediate consequence, consider a trajectory with constant velocity
after some fixed proper time $\tau_c$. Then for sufficiently late times 
\begin{eqnarray}
\tau-\tau_c\gg a^{-1} 
\end{eqnarray}
non-uniformities in velocity before
$\tau_c$ can be ignored and one can set
\begin{eqnarray}
\delta^{\pm}(\tau,\tau^\prime)=
D_{\beta}^{\pm}(\tau_c)\,(\tau-\tau^\prime),
\end{eqnarray}
which is equivalent to the stationary result (\ref{1.14}).

The Heisenberg field is constructed along the usual lines as
\begin{eqnarray}
\varphi(z)=\int_{0_+}^{\infty} \frac{d\omega}{(2\pi)2\omega}
\left(a_\omega^U\,U_\omega^R(z)+a_\omega^V\,V_\omega^R(z)+H.c.\right),
\label{1.17}
\end{eqnarray}
for points $z\equiv (u,v)$ in the right region, and similarly with 
$R\rightarrow L$ for $z$ in the left region. The commutations for the
creation-annihilation in operators are
\begin{eqnarray}
[a_\omega^U, a_{\omega^\prime}^{U+}]=[a_\omega^V, a_{\omega^\prime}^{V+}]=
2\pi(2\omega)\delta(\omega-\omega^\prime),
\label{1.18}
\end{eqnarray}
and zero in rest. To assure that canonical commutations are satisfied, we
shall suppose there exists a finite time $t_0$
(which can be taken arbitrarily far in the past) such that the trajectory
is uniform before $t_0$.
Then, in the reference frame where the mirror is at rest at infinite
past, the modes are given up to a certain moment by the $\beta=0$
coefficients (\ref{1.14}). Orthonormality and completeness can be
easily checked in this case. It follows that canonical commutations are
satisfied on some spacelike three surface, and thus, by Eq. (\ref{1.4})
they
are respected in whole Minkowski space \cite{bjorken}.

\section {The radiation}\label{three}

We are interested in the in-vacuum renormalized expectation values
$\langle T_{\mu\nu}\rangle_{ren}$ of the canonical energy-momentum operator
\begin{eqnarray}
T_{\mu\nu}=\partial_\mu\varphi\partial_\nu\varphi-
\frac{1}{2}g_{\mu\nu}g^{\alpha\beta}
\partial_\alpha\varphi\partial_\beta\varphi,
\label{op}
\end{eqnarray}
at points off the trajectory. We need the in-vacuum Wightman function
\begin{eqnarray}
D^+(z,z^\prime)=\langle 0, \mbox{in}\vert \phi(z) \phi(z^\prime) \vert 0,
\mbox{in}\rangle\!\,,
\label{2.1}
\end{eqnarray}
with the in vacuum 
defined by 
\begin{eqnarray}
a_\omega^U\vert 0, \mbox{in}\rangle\!=a_\omega^V\vert 0,
\mbox{in}\rangle\!=0,
\quad
\omega>0.
\label{1.19}
\end{eqnarray}

Let $D^+_0(z,z^\prime)$ denote the free field Wightman function, and let's 
introduce the renormalized value
\begin{eqnarray}
D^+_{ren}(z,z^\prime)=D^+(z,z^\prime)-D^+_0(z,z^\prime).
\label{2.2}
\end{eqnarray}
Then
\begin{eqnarray}
{\langle T_{uu}(u,v) \rangle}_{ren}&=&
\lim_{u^\prime\rightarrow u}
\lim_{v^\prime\rightarrow v}
\partial_u \partial_{u^\prime} D^+_{ren}(u,v,u^\prime,v^\prime),
\label{2.2*}\\
\langle T_{vv}(u,v)\rangle_{ren}&=&
\lim_{u^\prime\rightarrow u}
\lim_{v^\prime \rightarrow v}
\partial_v \partial_{v^\prime} D^+_{ren}(u,v,u^\prime,v^\prime).
\label{2.3}
\end{eqnarray}
Space time coordinates symmetrization \cite{birrell} proves unnecessary
here, as Eqs. (\ref{2.2*}), (\ref{2.3}) will lead to real results.
The mixed $uv$, $vu$ components vanish identically.

Let $D_{R}^+$ stand for $D^+(z,z^\prime)$ when both $z, z^\prime$ belong to 
$R$, and analogously for $D_{L}^+$. It is convenient to
introduce
$\tau_u$, $\tau_v$ as the inverse functions of $u$, $v$
\begin{eqnarray}
\tau_u(u(\tau))=\tau,
\quad
\tau_v(v(\tau))=\tau.
\end{eqnarray}
Simple calculations yield (note that $\omega$ dependence in
${\cal R}_\omega^{L,R}$ makes the frequency integrations
similar to that in the free field case)
\begin{eqnarray}
D^+_{R}=D_0^++D_{R1}^++D_{R2}^+,
\label{2.3**}
\end{eqnarray}
where ($\epsilon \rightarrow 0_+$)
\begin{eqnarray}
D_{R1}^+=\frac{a}{8\pi}
\int_{-\infty}^{\tau_u}d\tau_1
\,
\ln\, ((u(\tau_1)-u^\prime-i\epsilon)(v(\tau_1)-v^\prime-i\epsilon))\,
\nonumber\\
\nonumber\\
\times\exp \frac{a}{2}(\tau_1-\tau_u)\,
\hspace*{39pt}
\nonumber\\
\nonumber\\
+\frac{a}{8\pi}\int_{-\infty}^{\tau_{u^\prime}}d\tau_1
\,
\ln\, ((u-u(\tau_1)-i\epsilon)(v-v(\tau_1)-i\epsilon))
\nonumber\\
\nonumber\\
\times\exp \frac{a}{2}(\tau_1-\tau_{u^\prime})\,,
\end{eqnarray}
and
\begin{eqnarray}
D_{R2}^+= -\frac{a^2}{16\pi}
\int_{-\infty}^{\tau_u}
d\tau_1
\int_{-\infty}^{\tau_{u^\prime}}
d\tau_2
\ln\, ((u(\tau_1)-u(\tau_2)-i\epsilon)\hspace*{8pt}
\nonumber\\
\nonumber\\
\times (v(\tau_1)-v(\tau_2)-i\epsilon))\,
\exp \frac{a}{2}(\tau_1+\tau_2-\tau_u-\tau_{u^\prime}).
\label{2.6}
\nonumber\\
\end{eqnarray}
The explicit form of $D_0^+$ is of no relevance here.
$\tau_u$ is a shorthand for $\tau_u(u)$, and similarly for
$\tau_{u{^\prime}}$. $D^+_{L}$ can be obtained from $D_{R}^+$
with the substitutions
\begin{eqnarray}
u \leftrightarrow v, \quad u^\prime \leftrightarrow v^\prime, \quad
\tau_u \leftrightarrow \tau_v.
\label{parity}
\end{eqnarray}
We shall take advantage in the following of the $R-L$ formal symmetry above
to refer to $R$ quantities only.

Renormalization amounts to ignore $D_0^+$ in Eq. (\ref{2.3**}).
The double $v$ derivative (\ref{2.2*}) is trivial 
\begin{eqnarray}
\langle T_{vv}^R(u,v)\rangle _{ren}=0.
\label{2.8}
\end{eqnarray}
The $uu$ component (\ref{2.3}) requires some effort. Calculations are
outlined in Appendix A. One finds the $v$ independent quantity
\begin{eqnarray}
\langle T_{uu}^R(u,v)\rangle _{ren}=-\frac{1}{4\pi}
\frac{1-\beta^2(\tau_u)}{(1-\beta(\tau_u))^2}\times T^R(\tau_u),
\label{2.8*}
\end{eqnarray}
where
\begin{eqnarray}
T^R=T^R_{I}+T^R_{II},
\end{eqnarray}
with
\begin{eqnarray}
T^R_{I}(\tau_u)=\frac{a^2}{4}\int_{-\infty}^{\tau_u}\int_{-\infty}^{\tau_u}
d\tau_1 d\tau_2 \left[
\partial_{\tau_1}\partial_{\tau_2}\,\ln
\frac{v(\tau_1)-v(\tau_2)}{u(\tau_1)-u(\tau_2)}\right]
\nonumber\\
\nonumber\\
\times\exp \frac{a}{2}(\tau_1+\tau_2-2\tau_u),
\label{2.9}
\end{eqnarray}
and
\begin{eqnarray}
T^{R}_{II}(\tau_u)=-\frac{a}{2}\int_{-\infty}^{\tau_u}
\int_{-\infty}^{\tau_u} d\tau_1d\tau_2 \left[
\partial_{\tau_1}\partial_{\tau_2}
\frac{\dot u(\tau_1)-\dot u(\tau_2)}{u(\tau_1)-u(\tau_2)}\right]
\nonumber\\
\nonumber\\
\times\exp \frac{a}{2}(\tau_1+\tau_2-2\tau_u).
\label{2.10}
\end{eqnarray}
The overdot represents differentiation with respect to proper time.
Coincidence limits in the integrands are discussed in Appendix B.

Equations (\ref{2.8*})-(\ref{2.10}) represent our main result. They 
explicitly show that for finite barrier strengths the radiated 
energy-momentum density at a given point
\begin{eqnarray}
\langle T^R_{tt}(u,v)\rangle_{ren}=-\langle
T^R_{tx}(u,v)\rangle_{ren}\equiv
\langle T^R_{uu}(\tau_u)\rangle_{ren}
\end{eqnarray}
depends on the entire mirror's past history\footnote{
A similar conclusion was reached, e.g., in Refs. \cite {barton1,
jaekel1}.}.
Past dependence extends up to the retarded (emission) time $\tau_u$, 
marking the intersection of the trajectory with the past light cone at
$(u,v)$. As for ${\cal R}$, ${\cal T}$ coefficients, exponential factors 
restrict the influence of past motions to an effective interval $\sim a^{-1}$
before $\tau_u$.

Perfect reflectivity naturally results by letting 
$a\rightarrow \infty$.
To evaluate the limits, it is convenient to use
\begin{eqnarray}
\lim_{a\rightarrow\infty}a\int_{-\infty}^{\tau}
d\tau^\prime f(\tau^\prime)\exp a(\tau^\prime-\tau)
=\lim_{\stackrel{\tau^\prime\rightarrow \tau}{\tau^\prime<\tau}}
f(\tau^\prime),
\label{2.7}
\end{eqnarray}
which makes apparent disappearance of history dependence. For Eqs. (\ref{2.9}),
(\ref{2.10}) situation is as follows. Let $\alpha$ denote the mirror's
proper
acceleration. Then coincidence limit (\ref{lim1}) in $T^R_I$ can be 
rewritten (we assume the acceleration is continuously differentiable at
$\tau_u$, see Appendix B)
\begin{eqnarray}
\lim_{a\rightarrow \infty}
T^R_{I}(\tau_u)=\frac {1}{3}\dot \alpha(\tau_u).
\label{perfect}
\end{eqnarray}
For $T^R_{II}$ component, suffices to note that the double integral is
multiplied by only one power of $a$, so that it brings no
contribution. Combining Eqs. (\ref{perfect}), (\ref{2.8*}), one recovers the 
well known perfect mirror result \cite{fulling, davies}.

Suppose now that $a$ is finite, but sufficiently large that no
significative changes in velocity occur in the effective
interval before $\tau_u$. In other words, we are restricting to slow motions on
a proper time scale $\sim a^{-1}$. Then the brackets can be Taylor expanded
around $\tau_1=\tau_2=\tau_u$, followed by a term by term integration.
The result is the series
\begin{eqnarray}
T^R=T^R_0+\sum_{n=1}^\infty \left(\frac{2}{a}\right)^n T^R_n,
\label{n1}
\end{eqnarray}
with $T^R_{n}$ coefficients completely determined as polynomials in 
$\alpha$ and its first $n+1$ derivatives at $\tau_u$. $T^R_0$ equals precisely 
the perfect mirror contribution (\ref{perfect}). Higher terms naturally 
describe semitransparency corrections. We give below the
first two coefficients (for arbitrary orders, evaluation of the coincidence
limits in the Taylor expansion proves quite laborious)
\begin{eqnarray}
T^R_1&=&\frac{1}{6}\alpha \dot \alpha-\frac{1}{6}\ddot\alpha,
\label{n3} \\
T^R_2&=&\frac{37}{420}\alpha^2\dot \alpha
-\frac{1}{6}\dot \alpha^2
-\frac{1}{6}\ddot \alpha \alpha
+\frac{11}{120}\stackrel{...}{\alpha}.
\label{n4}
\end{eqnarray}
It is interesting to note that both corrections can be accommodated into
the perfect mirror result by replacing $\alpha$ in (\ref{perfect}) with an
``effective acceleration'' containing supplemental 
$\dot\alpha$, $\ddot\alpha$ dependent terms.

We also point out the following aspect: 
one sees that corrections can generally both diminish or enhance the zeroth
order flux. Consider, for example, a leftward accelerated trajectory on which the
perfect mirror radiates a positive energy flux in $R$
\begin{eqnarray}
\alpha <0, \quad \dot\alpha<0.
\end{eqnarray}
Then if the flux is increasing or not decreasing too fast as seen in the 
mirror proper frame (i.e. $\ddot\alpha$ is not too large), the leading 
correction Eq. (\ref{n3}) has always a diminishing effect.
By contrast, for the absolute values of the negative energy flux radiated
in $L$, the opposite is true.

We end with a comment on non-radiative motions. It is clear that 
non-radiativity results as a history dependent property. Consider a
fixed trajectory up to proper time $\tau_0$. To find its non-radiative 
future extension, one should solve the integro-differential
system
\begin{eqnarray}
T^R(\tau)=T^L(\tau)=0,\quad \tau>\tau_0,
\end{eqnarray}
where $\tau<\tau_0$ trajectory plays the role of Cauchy data. 
Unfortunately, a general treatment is analytically rather difficult, if not 
impossible, given non-linearities implied. 
We shall contend ourselves to point out the special case when both integrands 
in Eqs. (\ref{2.9}), (\ref{2.10}) vanish identically. Coincidence limit in 
$T^R_I$ readily gives a necessary condition as
\begin{eqnarray}
\alpha(\tau)=const.
\end{eqnarray}
One can easily check that this is also
sufficient. Hence no radiation occurs for constant 
acceleration trajectories, and this is what one should expect by 
conformal invariance of the theory\footnote{A similar situation happens 
for $D=2$ spherical mirrors in Ref. \cite{frolov}.}. An observation is in 
order here. It may be argued that uniformly accelerated trajectories do not 
respect the constant velocity condition for $t\rightarrow -\infty$ as formulated 
in Sec. \ref{two}. From a practical point of view, however, this is 
not essential: the mirror can be considered stationary up to a certain moment, 
and only afterwards taken to accelerate. After a proper time large enough 
compared to 
$a^{-1}$ the motion will effectively count as non-radiative.

\section{An analogy with extended classical charges}\label{five}

We conclude by briefly discussing an analogy. History dependence of the emitted 
flux invites to a parallel with extended charges in classical electrodynamics.
Consider an arbitrarily moving charge with characteristic length $r$.
Retardation implies that the electromagnetic flux at a certain point is 
determined by the motion in some finite
length interval $\Delta \tau\sim r$. One is led thus to the correspondence
\begin{eqnarray}
r \sim a^{-1}.
\label{ana}
\end{eqnarray}
It is further relevant to turn toward a dynamical aspect. 
It is a common remark that the quantum 
backreaction force experienced by perfect mirrors in two dimensions 
\cite{fulling} displays the same trajectory dependence
as the radiation force \cite{rohrlich} acting on pointlike
charges. This makes the well known $r\rightarrow 0$ pathologies, i.e. runaway 
trajectories and acausality, manifest too for perfect mirrors.

Our observation concerning similarity expressed by (\ref{ana}) is as follows. 
It was shown in Refs. \cite{jaekel3,jaekel4} 
that situation above changes when allowing for a transparency factor: more 
precisely, the unphysical behavior is absent when 
the transparency cut-off is sufficiently small compared to the mirror's 
mass\footnote{The result is based on the $S$ matrix approach in Ref. 
\cite{jaekel1}. There are certain limitations in the proof, as approximations 
are made to 
linearize the dynamical equation. In our model, linearization shows there are 
no unphysicalities if $a/(4\pi m)<1$, with $m$ the mirror's mass.}. 
Now, the cut-off in our model is some quantity of order $a$. 
These find a direct parallel in what happens for 
extended charges: the inverse of $r$ plays an identical role, in the sense 
that unphysicalities disappear \cite{franca} provided the ratio between 
the electromagnetic self energy $\sim r^{-1}$ and the physical mass is 
small enough. 

\bigskip
\noindent
\section*{Acknowledgments}
\bigskip
I thank to Farkas Atilla for stimulating discussions and
for revising the manuscript. I am grateful to Claudia Eberlein for putting
me in contact with her results. I also thank to Hanna Csunderlik for useful 
observations and criticisms.

\begin{center}
\section*{Appendix A}
\end{center}
We sketch here the steps leading to Eqs. (\ref{2.9}),
(\ref{2.10}). Let
\begin{eqnarray}
t_{1,2}(u,u^\prime)=\partial_u\partial_{u^\prime}D_{R1,2}^+(u,v,u^\prime,
v^\prime).
\end{eqnarray}
It turns out that $t_1$, $t_2$ separately diverge for $\epsilon\rightarrow
0$,
$u^\prime \rightarrow u$. An appropriate grouping of terms in their
sum, however, makes divergences to cancel among
themselves. We show below a way to do this. Observe that terms in
$t_2$ can be organized in $u$, $v$ dependent parts. We denote them as 
$t_2^u$, $t_2^v$, respectively. We further introduce a set of equivalent, 
but $formally$ different expressions \begin{eqnarray}
t_{2\alpha}^u=t_{2\beta}^u=t_{2\gamma}^u=t_2^u,
\quad
t_{2\gamma}^v=t_2^v,
\end{eqnarray}
which we define as follows.

With a series of integrations by parts with respect to
$\tau_1$, $t_2^u$ can be cast into the form
\begin{eqnarray}
t_{2\alpha}^{u}=-\frac{a^2}{16\pi}\frac {d\tau_u}{du}
\frac{d\tau_{u^\prime}}{du^\prime}
\biggl\lbrace
\int_{-\infty}^{\tau_u}d\tau_1 \frac{du_1}{d\tau_1}
\frac{\exp\,\frac{a}{2}(\tau_1-\tau_u)}{u(\tau_1)-u^\prime-i\epsilon}
\nonumber\\
-\frac{a}{2}\int_{-\infty}^{\tau_u}d\tau_1 \frac{du_1}{d\tau_1}
\int_{-\infty}^{\tau_{u^\prime}}
d\tau_2
\frac{\exp\,\frac{a}{2}(\tau_1+\tau_2-\tau_u-\tau_{u^\prime})}
{u(\tau_1)-u(\tau_2)-
i\epsilon}
\biggl\rbrace.\label{A.1}
\end{eqnarray}
Similar integrations with respect to $\tau_2$ lead to a second expression
$t_{2\beta}^u$. An integration by parts with respect to $\tau_2$ in the
double integral 
in Eq. (\ref{A.1}) eliminates the single integration term. Let
$t_{2\gamma}^u$ denote the resulting expression. Starting with $t_2^v$ and
repeating the steps leading from $t_2^u$ to $t_{2\gamma}^u$, one obtains
$t_{2\gamma}^v$.

We organize now the sum $t_1+t_2$ as
\begin{eqnarray}
(t_{2\gamma}^v-t_{2\gamma}^u)+(t_{2\alpha}^u+t_{2\beta}^u+t_1).
\label{para}
\end{eqnarray}
At this point one can set $\epsilon=0$ and let $u^\prime \rightarrow u$, in
each parenthesis the divergences getting cancelled in a manifest way. The
two parentheses correspond, in order, to $I$ and $II$ contributions in
$\langle T_{uu}^R(u,v)\rangle _{ren}$.

\begin{center}
\section*{Appendix B}
\end{center}
We make here some observations concerning the coincidence limits in
Eqs. (\ref{2.9}), (\ref{2.10}). Simple analysis shows that the quantities
in
the brackets are finite, provided the trajectory is at least of class
$C^3$,
respectively $C^4$. The coincidence limit
in $T_I^{R}$ is of direct relevance for the perfect reflectivity limit.
For $\tau_{1,2}\rightarrow \tau$ it reads
\begin{eqnarray}
\frac{1}{6}\frac{\stackrel{...}{v}(\tau)}{\dot v(\tau)}-
\frac{1}{6}\frac{\stackrel{...}{u}(\tau)}{\dot u(\tau)}.
\label{lim1}
\end{eqnarray}

An equivalent way of writing $T^R_{I,II}$ is to eliminate the double
derivatives
$\partial_{\tau_1}\partial_{\tau_2}$ by integrations by
parts. Then coincidence limits involve only
the first derivatives of $u$, $v$ in $T_{I}^{R}$, and the first two
derivatives
of $u$ in $T_{II}^{R}$. The same applies for the resulting boundary terms
at
$\tau_u$. The conclusion is that a $C^2$ trajectory suffices to 
assure the continuity
of the flux. Note that this is $not$ the case for perfect mirrors.

With the above choice, discontinuities in the acceleration manifest in 
$T_{II}^{R}$ in the boundary term at $\tau_u$. Consider a trajectory with
acceleration continuous everywhere excepting the moment $\tau_0$, where
\begin{eqnarray}
\ddot u(\tau_{0+})- \ddot u(\tau_{0-})=\Delta \ddot u_0\neq 0,
\quad \tau_{0\pm}=\tau_0+0_\pm.
\end{eqnarray}
One finds this entails the flux discontinuity
\begin{eqnarray}
\langle T^R_{uu}(u(\tau_{0+}),v)\rangle_{ren}-
\langle T^R_{uu}(u(\tau_{0-}),v)\rangle_{ren}
=\frac{a}{8\pi}\frac{\Delta \ddot u_0}{\dot u^3(\tau_0)}.
\label{disco}
\end{eqnarray}
As $a$ approaches infinity, the quantity above is progressively canceled by
the integration terms in $T_{II}^{R}$. In the limit, however,
the discontinuity survives due to the $T_{I}^R$ contribution (when it
assumes a distributional form cf. (\ref{lim1})).

\end{document}